\documentclass[aip,amsmath,amssymb,reprint]{revtex4-1}

\usepackage{algorithm}
\usepackage{graphicx}
\usepackage{dcolumn}
\usepackage{bm}

\usepackage{multirow}
\usepackage{array}
\newcolumntype{P}[1]{>{\centering\arraybackslash}p{#1}}
\usepackage{longtable}

\usepackage{makecell}
\usepackage{listings}
\usepackage{multirow}
\usepackage{siunitx}
\usepackage{xcolor}

\usepackage{afterpage}
\usepackage{float}

\def\beq{\begin{eqnarray}}
\def\eeq{\end{eqnarray}}
\def\V{\mathcal{V}}

\def\P{\mathcal{P}}

\def\brakett#1#2#3{\langle #1 \vert #2 \vert #3 \rangle}

\usepackage{hyperref}

\begin{document}

\title{Almost exact energies for the Gaussian-2 set with the semistochastic heat-bath configuration interaction method}

\author{Yuan Yao}
  \email{yy682@cornell.edu}
  \affiliation{Laboratory of Atomic and Solid State Physics, Cornell University, Ithaca, New York 14853, United States}
\author{Emmanuel Giner}
  \email{emmanuel.giner@lct.jussieu.fr}
  \affiliation{Laboratoire de Chimie Th\'eorique, Sorbonne Universit\'e and CNRS, F-75005 Paris, France}
\author{Junhao Li}
  \email{jl2922@cornell.edu}
  \affiliation{Laboratory of Atomic and Solid State Physics, Cornell University, Ithaca, New York 14853, United States}
\author{Julien Toulouse}
  \email{toulouse@lct.jussieu.fr}
  \affiliation{Laboratoire de Chimie Th\'eorique, Sorbonne Universit\'e and CNRS, F-75005 Paris, France}
  \affiliation{Institut Universitaire de France, F-75005 Paris, France}
\author{C. J. Umrigar}%
  \email{cyrusumrigar@cornell.edu}
  \affiliation{Laboratory of Atomic and Solid State Physics, Cornell University, Ithaca, New York 14853, United States}

\date{\today}

\begin{abstract}
The recently developed semistochastic heat-bath configuration interaction (SHCI) method is a systematically
improvable selected configuration interaction plus perturbation theory method capable of giving essentially exact
energies for larger systems than is possible with other such methods.
We compute SHCI atomization energies for 55 molecules which have been used as a test set in prior studies
because their atomization energies are known from experiment.
Basis sets from cc-pVDZ to cc-pV5Z are used, totaling up to 500 orbitals and a Hilbert
space of $10^{32}$ Slater determinants for the largest molecules. For each basis, an extrapolated
energy well within chemical accuracy (1 kcal/mol or 1.6 mHa/mol) of the exact energy for that basis is computed using only a tiny
fraction of the entire Hilbert space. We also use our almost exact energies to benchmark coupled-cluster 
[CCSD(T)] energies. The energies are extrapolated to the complete basis set limit and compared to the experimental atomization
energies.
The extrapolations are done both without and with a basis-set correction based on density-functional theory.
The mean absolute deviations from experiment for these extrapolations are
0.46 kcal/mol and 0.51 kcal/mol, respectively.
Orbital optimization methods used to obtain improved convergence of the SHCI energies are also discussed. 

\end{abstract}

\maketitle

\section{Introduction}
The recently developed semistochastic heat-bath configuration interaction (SHCI)
method~\cite{HolTubUmr-JCTC-16,ShaHolJeaAlaUmr-JCTC-17,HolUmrSha-JCP-17,SmiMusHolSha-JCTC-17,MusSha-JCTC-18,ChiHolOttUmrShaZim-JPCA-18,LiOttHolShaUmr-JCP-18}
is a systematically improvable quantum chemistry method capable of providing essentially exact energies for small
many-electron systems.
It has been successfully applied to a number of challenging problems in quantum chemistry, including the potential
energy curve of the chromium dimer~\cite{LiYaoHolOttShaUmr-PRR-20} for which coupled cluster with single, double, and
perturbative triple excitations [CCSD(T)], the gold standard of single-reference quantum chemistry, does not
give even a qualitatively correct description.
It has also been used as the reference method for calculations on transition metal atoms, ions,
and monoxides~\cite{WilYao_etal_UmrWag-PRX-20} to test the accuracy of a wide variety of other electronic-structure methods.

SHCI is an example of the selected configuration interaction (SCI) plus perturbation theory (SCI+PT)
methods~\cite{BenDav-PR-69,WhiHac-JCP-69,HurMalRan-JCP-73,BuePey-TCA-74,EvaDauMal-CP-83,GinSceCaf-CJC-13,Eva-JCP-14,SceAppGinCaf-JCoC-16,GarSceLooCaf-JCP-17,LooSceBloGarCafJac-JCP-18,HaiTubLevWhaHea-JCTC-19,LooLipPasSceJac-JCTC-20} which have two stages.
In the first stage a variational wave function is constructed iteratively, starting from
a determinant that is expected to have a significant amplitude in the final wave function, e.g., the Hartree-Fock (HF)
determinant.  The number of determinants in the variational wave function is controlled by a parameter $\epsilon_1$.
In the second stage, second-order perturbation theory is used to improve upon the variational energy. The
total energy (sum of the variational energy and the perturbative correction) is computed at several values of
$\epsilon_1$ and extrapolated to $\epsilon_1 \to 0$ to obtain an estimate for the full configuration interaction (FCI) energy. The efficiency of SHCI depends on the choice of the orbitals -- natural orbitals lead to faster convergence of
the energy relative to HF orbitals and optimized orbitals yield yet faster convergence.

In this paper, 
the SHCI method is reviewed in Section~\ref{SHCI_review}, our orbital optimization schemes are described
in Section~\ref{Orb_opt}, the basis-set correction and extrapolation that we use are discussed in Section~\ref{basis_set}, and the details of the calculations are given in
Section~\ref{Comput_details}.
In Section~\ref{Results} we apply SHCI to the 55 first- and second-row molecules that served as the training set for the Gaussian-2 (G2) protocol~\cite{CurRagTruPop-JCP-91}
because accurate experimental atomization energies were believed to be known for them.
The G2 protocol is one of several quantum chemistry composite methods that combine low-order methods on
large basis sets and high-order coupled-cluster methods on smaller basis sets to compute accurate thermochemical
properties (see, e.g., Refs.~\onlinecite{CurRedRag-JCP-07,FelPet-JCP-99,TajSzaCsaKalGauValFloQueSta-JCP-04,KarRabMarRus-JCP-06,ThoSta_etal-JCP-19}.).
These 55 molecules, which we refer to as the G2 set, have previously been used to test the accuracy of coupled-cluster-based
methods~\cite{FelPet-JCP-99} and quantum Monte Carlo (QMC) methods~\cite{Gro-JCP-02,NemTowNee-JCP-10,PetTouUmr-JCP-12,CafAppGinSce-ACS-16}.
We employ the correlation consistent basis sets cc-pV$n$Z for $n$ = 2 (D), 3 (T), 4 (Q),
and 5~\cite{Dun-JCP-89}, keeping the core electrons frozen, to obtain SHCI energies that we believe are well within 1 mHa of the
exact (FCI) energies for each of the molecules and basis sets.
Hence these calculations provide a set of reference energies that can be used to test other
accurate electronic-structure methods.

The molecules in the G2 set are sufficiently weakly correlated that one would expect CCSD(T) to be reasonably
accurate, but not at the level of 1 mHa.  Hence, we calculate also the CCSD(T) energies using the same basis sets
in order to use SHCI  to evaluate the errors in the CCSD(T) energies, as FCI is not feasible for most of these systems.
The SHCI energies are then extrapolated to the complete-basis-set (CBS) limit, both without and with a basis-set correction based on density-functional theory (DFT)~\cite{GinPraFerAssSavTou-JCP-18,LooPraSceTouGin-JPCL-19,GinSceTouLoo-JCP-19,GinSceLooTou-JCP-20}.
Corrections taken from the literature for zero-point energy, relativistic effects, and core-valence correlation are
then applied to obtain our predictions for the atomization energies, which are then compared to the best available
experimental values.  For some systems the available experimental values differ substantially from each other and for
at least one system we believe that the theoretical estimates are more accurate than the best experimental value.

\section{Review of the SHCI method}
\label{SHCI_review}
In this section, we review the SHCI method,
emphasizing the two important ways it differs from other SCI+PT methods.
In the following, we use $\V$ for the set of variational determinants, and $\P$ for the set of perturbative
determinants, that is, the set of determinants that are connected to the variational determinants by at least one
non-zero Hamiltonian matrix element but are not present in $\V$.

\subsection{Variational stage}
\label{Var}

SHCI starts from an initial determinant
and generates the variational wave function through an iterative process.
At each iteration, the variational wave function, $\Psi_V$, is written as a linear combination of the determinants
in the space $\V$
\begin{align}
\left|\Psi_{V} \right\rangle= \sum_{D_i \in \V} c_{i} \left|D_{i}\right\rangle
\end{align}
and new determinants, ${D_a}$, from the space $\P$ that satisfy the criterion
\beq
\exists\; D_i \in \V , \mathrm{\ such\ that\ } \left|H_{a i} c_{i}\right| \ge \epsilon_{1}
\label{HCI_criterion}
\eeq
are added to the $\V$ space, where
$H_{ai}$ is the Hamiltonian matrix element between determinants $D_a$ and $D_i$, and
$\epsilon_1$ is a user-defined parameter that controls the accuracy of the variational
stage~\footnote{Since the absolute values of $c_i$ for the most important determinants tends to go down as more
determinants are
included in the wave function, a somewhat better selection of determinants is obtained by using a larger value of
$\epsilon_1$ in the initial iterations.}.
(When $\epsilon_1=0$, the method becomes equivalent to FCI.)
After adding the new determinants to $\V$, the Hamiltonian matrix is constructed and diagonalized using the diagonally
preconditioned Davidson method~\cite{Dav-CPC-89} to obtain an improved estimate of the lowest eigenvalue, $E_{V}$,
and eigenvector, $\Psi_V$.
This process is repeated until the change in the variational energy $E_V$ falls below a certain threshold.

Other SCI methods use different criteria, based on
either the first-order perturbative coefficient of the wave function,
\beq
\left|c_a^{(1)}\right|=\left|\frac{\sum_i H_{ai}c_i}{E_V-E_a}\right| > \epsilon_1
\label{eq:cipsi_wf}
\eeq
or the second-order perturbative correction to the energy,
\beq
-\Delta E^{(2)}=-\frac{\left(\sum_i H_{ai}c_i\right)^2}{E_V-E_a} > \epsilon_1,
\label{eq:cipsi_energy}
\eeq
where $E_a = H_{aa}$.
The reason we choose instead the selection criterion in Eq.~(\ref{HCI_criterion}) is that it can be implemented
very efficiently without checking the vast majority of the determinants that do not meet the criterion, by taking
advantage
of the fact that
most of the Hamiltonian matrix elements correspond to double excitations, and their values do not depend
on the determinants themselves but only on the four orbitals whose occupancies change during the double excitation.
Therefore, at the beginning of an SHCI calculation, for each pair of spin-orbitals, the absolute values of the Hamiltonian
matrix elements obtained by doubly exciting from that pair of orbitals is computed and stored
in decreasing order by magnitude, along with the corresponding pairs of orbitals the electrons would excite to.
Then the double excitations that meet the criterion in Eq.~(\ref{HCI_criterion}) can be generated by
looping over all pairs of occupied orbitals in the reference determinant, and
traversing the array of sorted double-excitation matrix elements for each pair.
As soon as the cutoff is reached, the loop for that pair of occupied orbitals is exited.
Although the criterion in Eq.~(\ref{HCI_criterion}) does not include information from the diagonal elements,
this selection criterion is not significantly different from either of the criteria
in Eqs.~(\ref{eq:cipsi_wf}) and (\ref{eq:cipsi_energy}) because
the terms in the numerators of Eqs.~(\ref{eq:cipsi_wf}) and (\ref{eq:cipsi_energy})
span many orders of magnitude, so the sums are highly correlated with the largest-magnitude term in the sums in
Eqs.~(\ref{eq:cipsi_wf}) or (\ref{eq:cipsi_energy}), and because the denominator is never small
after several determinants have been included in $\V$.
It was demonstrated in Ref.~\onlinecite{HolTubUmr-JCTC-16} that the selected determinants give only slightly
inferior convergence
to those selected using the criterion in Eq.~(\ref{eq:cipsi_wf}).  This is greatly outweighed by the improved
selection speed.
Moreover, one could use the criterion in Eq.~(\ref{HCI_criterion}) with a smaller value of $\epsilon_1$ as a
preselection criterion, and then select determinants
using the criterion in Eq.~(\ref{eq:cipsi_energy}) or something close to it, thereby having the benefit of both a
fast selection method and a
close to optimal choice of determinants.
We use a similar, but somewhat more complicated criterion, also for the selection of the determinants
connected to those in $\V$ by a single excitation, but this improvement is of lesser importance
because the number of determinants connected by single excitations is much smaller than the number connected by double excitations.
With these improvements the time required for selecting determinants is negligible, and
the most time consuming step by far in the variational stage is the construction of the sparse
Hamiltonian matrix.  Details for doing this efficiently are given in Ref.~\onlinecite{LiOttHolShaUmr-JCP-18}.

\subsection{Perturbative stage}
\label{PT}

In common with most other SCI+PT methods, the perturbative correction is
computed using Epstein-Nesbet perturbation theory~\cite{Eps-PR-26,Nes-PRS-55}.
The variational wave function is used to define the zeroth-order Hamiltonian, $\hat{H}^{(0)}$, and the perturbation, $\hat{H}^{(1)}$,
\begin{align}
	\hat{H}^{(0)} &= \sum_{D_i,D_j \in \V} H_{ij} |D_i\rangle\langle D_j| + \sum_{D_a \notin \V } H_{aa} |D_a\rangle\langle
	D_a|. \nonumber\\
	\hat{H}^{(1)} &= \hat{H} - \hat{H}^{(0)} . \label{eq:part}
\end{align}
The first-order energy correction is zero, and the second-order energy correction $\Delta E^{(2)}$ is
\beq
 \Delta E^{(2)} &=& \langle\Psi_V|\hat{H}^{(1)}|\Psi^{(1)}\rangle
 \;=\; \sum_{D_a \in \P} \frac{\left(\sum_{D_i \in \V} H_{ai} c_i\right)^2}{E_V - E_a},
\label{eq:PTa}
\eeq
where $\Psi^{(1)}$ is the first-order wave-function correction.
The SHCI total energy is
\beq
\label{eq:E_tot}
E^{\rm SHCI} &=& E_V + \Delta E^{(2)} \;=\; \brakett{\Psi_V}{H}{\Psi_V} + \Delta E^{(2)}
\eeq

It is expensive to evaluate the expression in Eq.~(\ref{eq:PTa}) because the outer summation includes all determinants
in the space $\P$ and their number is
${\cal O}(N_\text{e}^2 N_\text{v}^2 N_\V)$, where $N_\V$ is the number of variational determinants, $N_\text{e}$ is the number of electrons,
and $N_\text{v}$ is
the number of unoccupied orbitals.
The straightforward
and time-efficient approach to computing the perturbative correction requires storing
the partial sum $\sum_{D_i \in \V} H_{ai} c_i$ for each unique $a$, while
looping over all the determinants $D_i \in \V$. This creates a severe memory bottleneck.
An alternative approach, which is widely used, does not require storing the unique $a$, but requires checking whether the determinant was
already generated by checking its connection with variational determinants whose connections have already been included.
This entails some additional computational expense.

The SHCI algorithm instead uses two other strategies to reduce both the computational time and the storage requirement.
First, SHCI screens the sum~\cite{HolTubUmr-JCTC-16} using a second threshold, $\epsilon_2$ (where
$\epsilon_2<\epsilon_1$) as the criterion for selecting perturbative determinants $D_a \in \P$,
\begin{equation}
\Delta E^{(2)} \left(\epsilon_{2}\right) = \sum_{D_a \in \P} \frac{\left(\sum_{D_i \in \V}^{(\epsilon_{2})}	H_{a i} c_{i}\right)
^{2}}{E_{V} - E_a}
\label{eq:PTb}
\end{equation}
where $\sum^{(\epsilon_{2})}$ indicates that only terms in the sum for which $\left|H_{a i} c_{i}\right| \ge
\epsilon_{2}$ are included.
Similar to the variational stage, we find the connected determinants efficiently with precomputed arrays of
double excitations sorted by the magnitude of their Hamiltonian matrix elements~\cite{HolTubUmr-JCTC-16}.
Note that the vast number of terms that do not meet this criterion are \emph{never evaluated}.

\begin{figure}
  \includegraphics[width=0.48\textwidth]{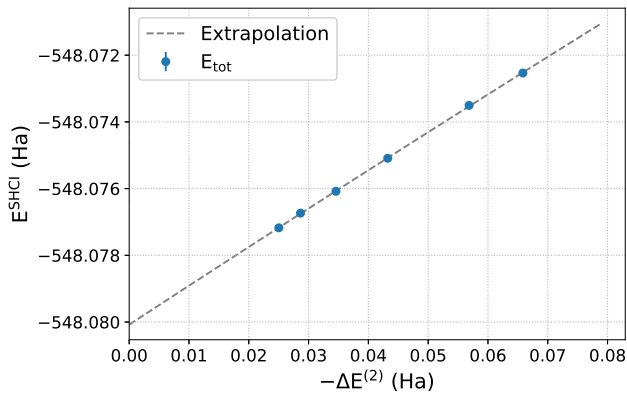}
  \caption{Convergence of SHCI energy of SO$_2$ in the cc-pV5Z basis set.  The line is a weighted quadratic fit, but
  is very nearly linear.  The statistical error bars are plotted but are invisible on the scale of the plot.}
  \label{fig:SHCI_extrap}
\end{figure}

Even with this screening, the simultaneous storage of all terms indexed by $a$ in Eq.~(\ref{eq:PTb}) can exceed
computer memory
when $\epsilon_2$ is chosen small enough to obtain essentially the exact perturbation energy.
The second innovation in the calculation of the SHCI perturbative correction is to overcome this memory bottleneck
by evaluating it semistochastically.
The most important contributions are evaluated deterministically and the rest are sampled stochastically.
Our original method used a two-step perturbative algorithm~\cite{ShaHolJeaAlaUmr-JCTC-17}, but our
later three-step perturbative algorithm~\cite{LiOttHolShaUmr-JCP-18} is even more efficient.
The three steps are:
\begin{enumerate}
  \item A deterministic step with cutoff $\epsilon_2^{\rm dtm} (< \epsilon_1)$, wherein
  all the variational determinants are used, and
  all the perturbative batches are summed over.
  \item A ``pseudo-stochastic" step, with cutoff $\epsilon_2^{\rm psto} (< \epsilon_2^{\rm dtm})$, wherein
  all the variational determinants are used, but the perturbative determinants are partitioned
        into batches.
  Typically only a small fraction of these batches
  need be summed over to achieve an error much smaller than the target error.
  \item A stochastic step, with cutoff $\epsilon_2 (<\epsilon_2^{\rm psto}) $, wherein a few stochastic
  samples of variational determinants,
  each consisting of $N_d$ determinants, are sampled with probability $\vert c_i \vert/\sum_{D_i \in \V} \vert c_i \vert$,
  and only one of the perturbative batches is randomly selected per variational sample.
\end{enumerate}
Using this semistochastic algorithm, the statistical error of our calculations for each $\epsilon_1$ is at most 20 $\mu$Ha,
which is negligible on the scale of the desired accuracy.
Having a small statistical error is important for doing a reliable extrapolation to the $\epsilon_1=0$ limit.
This is done~\cite{HolUmrSha-JCP-17} by computing $E^{\rm SHCI}$ at 5 or 6 values of $\epsilon_1$ and using a weighted quadratic fit of
$E^{\rm SHCI}$ to $-\Delta E^{(2)}$ to obtain $E^{\rm SHCI}$ at $-\Delta E^{(2)}=0$, using weights proportional to $(\Delta E^{(2)})^{-2}$.
Fig.~\ref{fig:SHCI_extrap} shows the convergence of $E^{\rm SHCI}$ for the system that has the
largest extrapolation distance (difference between the energy at the smallest $\epsilon_1$ used and the estimated energy at $\epsilon_1=0$),
namely, SO$_2$ in the cc-pV5Z basis set.

We note that, subsequent to our first semistochastic paper~\cite{ShaHolJeaAlaUmr-JCTC-17}, a completely different, but also efficient, semistochastic approach
has been presented in Ref.~\onlinecite{GarSceLooCaf-JCP-17}.

\section{Orbital optimization} \label{Orb_opt}

SHCI gives an estimate of the exact FCI energy by extrapolating energies evaluated at several
$\epsilon_{1}>0$ to $\epsilon_{1} =0$, the FCI limit.
This results in an extrapolation error that disappears in the limit that the extrapolation
distance 
goes to zero.

The extrapolation distance can be reduced by decreasing $\epsilon_{1}$, but this is limited by
the available computer memory and time.
An alternative approach is to optimize the
orbitals to obtain more compact configuration-interaction (CI) expansions with lower variational energies.

\begin{figure}
  \includegraphics[width=0.5\textwidth]{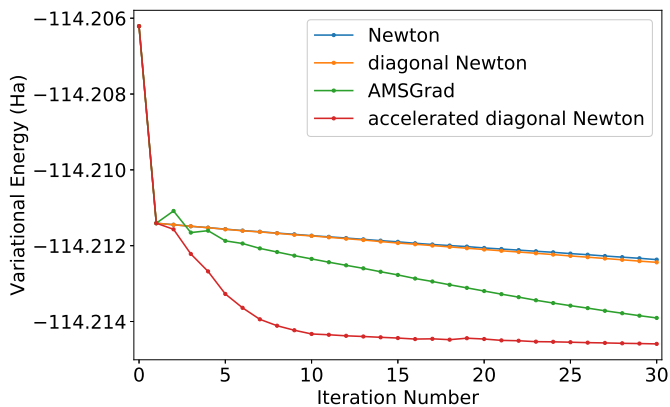}
  \caption{Comparison of four orbital optimization schemes for the H$_2$CO molecule in the cc-pVDZ basis and threshold parameter $\epsilon_1 = 2 \times 10^{-4}$. All four calculations start with HF orbitals and construct natural orbitals on the first iteration, so they differ only from the second iteration on.
  The Newton and diagonal Newton curves are nearly coincident for this system.}
  \label{fig:optimization}
\end{figure}

The first step to orbital optimization is to find the SHCI natural orbitals, i.e., the eigenstates of
the one-body reduced density matrix. These orbitals have a definite occupation number for a given variational
wave function and the most occupied ones represent in some sense the most important degrees of freedom.

Orbitals can be further optimized by directly minimizing the energy of the variational wave function through the
orbital rotation parameters ${\bf X}$:
\begin{align}
\label{energy}
E({\bf X}) = \langle \Psi_V | \exp(\hat{X}) \hat{H} \exp(-\hat{X}) | \Psi_V \rangle,
\end{align}
where $\hat{X}$ is a real anti-Hermitian operator such that $\exp(-\hat{X})$ parameterizes unitary
transformations in orbital space.
For a system with $N_{\rm orb}$ real-valued orbitals, this yields at most $N_{\rm orb}(N_{\rm orb}-1)/2$ orbital optimization parameters, which are the
elements of the real antisymmetric matrix ${\bf X}$.
In reality, the number of parameters will often be less than this due to point-group symmetry.
Depending on the particular optimization algorithm used, the gradient and sometimes part of the Hessian of the energy with respect to the orbital parameters are needed, either of which requires computing both the one- and two-body density matrices of the variational wave function.
In addition to the orbital parameters, the CI parameters (which are much more numerous) must
be optimized as well.
We next discuss some of the optimization methods we have studied.

\subsection{Newton's method}
Newton's method is a straightforward method for optimizing the parameters.
The parameters ${\bf x}_{t+1}$ at iteration $t+1$ are given by
\begin{equation}
{\bf x}_{t+1} = {\bf x}_t - {\bf h}_t^{-1}{\bf g}_t .
\end{equation}
where ${\bf g}_t$ and ${\bf h}_t$ are the gradient and the Hessian of the energy with respect to the parameters at iteration $t$.
In practice it is more efficient to find the parameter changes by solving the set of linear equations:
\beq
{\bf h}_t \left({\bf x}_{t+1} - {\bf x}_t \right) &=& -{\bf g}_t .
\eeq
However, the problem is that the number of parameters is typically much too large for even this to be
practical.  Typically, even using a rather large value of the threshold parameter $\epsilon_1$ for the optimization step, there
are millions of CI parameters whereas there are only thousands of orbital parameters.
So, one resorts to alternating the optimization of the CI parameters using the usual Davidson algorithm,
and optimizing the orbital parameters in the much smaller space of orbital rotations using Newton's method.
This alternating optimization often converges very slowly because the coupling between the CI parameters
and the orbital parameters is strong as can be seen in Fig.~\ref{fig:optimization}.
Note that the orbital optimization problem in SHCI is more difficult than that in the usual
complete-active-space self-consistent-field (CASSCF) method for two reasons.
First, none of the orbital rotations among orbitals of the same symmetry are redundant, so the number
of orbital parameters that need to be optimized is much larger.
Second, the coupling between the CI parameters and the orbital parameters is stronger.

In quantum chemistry problems, the orbital part of the Hessian matrix is often diagonally dominant.
In that case one can save significant computer time by ignoring the off-diagonal elements.
We refer to this as the ``diagonal Newton" method, and Fig.~\ref{fig:optimization} shows that for this
molecule it converges at the same rate as Newton's method.  The convergence of both methods is
limited by the lack of coupling between the CI and orbital parameters.

\subsection{AMSGrad}
AMSGrad is a momentum-based gradient-descent method commonly used in machine learning~\cite{RedKalKum-ICLR-18}. It avoids the expensive
Hessian calculations since only gradient information is needed.
At each iteration, it employs running averages of the gradient components and their squares, determined
by the mixing parameters $\beta_1, \beta_2 \in (0,1)$, according to
\begin{align}
{\bf m}_t &= \beta_1 {\bf m}_{t-1}+(1-\beta_1){\bf g}_t, \nonumber\\
v_t &= \beta_2 v_{t-1}+(1-\beta_2){\bf g}_t^2, \nonumber\\
\hat{v}_t &= \max(\hat{v}_{t-1}, v_t), \nonumber\\
{\bf x}_{t+1} &= {\bf x}_{t} - \frac{\eta}{\sqrt{\hat{v_t}} + \epsilon} {\bf m}_t.
\end{align}
The learning parameters $\eta, \beta_1,$ and $\beta_2$ together determine the level of aggressiveness of the
descent and $\epsilon$ is a small constant for numerical stability. We have found empirically that with a suitable level of aggressiveness, AMSGrad oscillates for the
first few iterations but eventually descends at a much quicker pace per iteration compared to either Newton or diagonal Newton,
as can be seen in Fig. \ref{fig:optimization}.
In addition each iteration takes less time since only the gradient is needed.
For a variety of systems we have found that the parameters $\eta=0.01, \beta_1=0.5,\beta_2=0.5$
give reasonably good convergence, even though they are much different from the values recommended in
the literature.

\subsection{Accelerated Newton's method}
Finally, we have developed a heuristic overshooting method that achieves yet better convergence for
most systems. Here, the overshooting tries to account for the coupling between CI and orbital parameters,
but it may be more generally useful whenever alternating optimization of subsets of parameters is done.

At each iteration, a diagonal Newton step is calculated for the orbital parameters, but, instead of using the proposed step, it is amplified by
a factor $f_t$ determined by the cosine of the angle between the previous step ${\bf x}_{t}-{\bf x}_{t-1}$ and the current step ${\bf x}_{t+1}-{\bf x}_{t}$:
\begin{align}
    f_t=\min \left( \frac{1}{2-\cos({\bf x}_{t}-{\bf x}_{t-1},{\bf x}_{t+1}-{\bf x}_{t})},\frac{1}{\epsilon}\right)
\end{align}
where $\epsilon$ is initialized to $0.01$ and $\epsilon\leftarrow \epsilon^{0.8}$ each time
$\cos({\bf x}_{t}-{\bf x}_{t-1},{\bf x}_{t+1}-{\bf x}_{t}) <0$.
The cosine in the expression is calculated in a ``scale-invariant" way to make it invariant under
a rescaling of some of the parameters,
i.e., in the usual definition $\cos({\bf v},{\bf w})=\langle{\bf v},{\bf w}\rangle/\sqrt{\langle{\bf v},{\bf v}\rangle \langle{\bf w},{\bf w}\rangle}$
we define the inner product as $\langle{\bf v},{\bf w}\rangle = {\bf v}^T {\bf h} {\bf w}$,
where the Hessian $\bf h$ can again be approximated by its diagonal.
Another scale invariant choice for the inner product is
$\langle{\bf v},{\bf w}\rangle = {\bf v}^T {\bf g} {\bf g}^T {\bf w}$, and that works equally well.
 
As shown in Fig. \ref{fig:optimization}, this accelerated scheme optimizes much faster than the previous schemes. For instance, after 4 iterations, the gain in variational energy is already better than that after 20 iterations using
the conventional Newton's method. Compared to AMSGrad, the higher per iteration
cost is more than made up by the greatly reduced number of iterations needed.
For this system, not only does the energy drop significantly but the number of determinants decreases as well.
For the accelerated scheme the drop is from 145,370 to 93,882 determinants.
However, for some systems the number of determinants increases, thereby partly offsetting the benefit of the energy gain.

\section{Basis-set correction and extrapolation}
\label{basis_set}

We employ the correlation consistent polarized valence (cc-pV$n$Z) basis sets with $n=$ 2 (D), 3 (T), 4 (Q), 5.
The energies computed for each atom or molecule are extrapolated to the CBS limit
using separate extrapolations for the HF energy and the correlation energy,~\cite{HelKloKocNog-JCP-97,HalHelJorKloKocOlsWil-CPL-98,HalHelJorKloOls-CPL-99}
\beq
\label{HF_extrap}
    E^{\rm HF}_{\rm CBS} &=& E^{\rm HF}_n + a \exp{(-bn)}, \\
\label{corr_extrap}
    E^{\rm corr}_{\rm CBS} &=& E^{\rm corr}_n + c n^{-3}.
\eeq
where $n$ is the cardinal number of the basis set.
The only exception is Li, for which the lowest HF energy is taken as the CBS energy because the energies for $n=3,4,5$ cannot be fit by a decaying exponential.
Note that the correlation energy extrapolation has 2 parameters, so it is necessary to use only the $n=4$ and $5$ basis sets, whereas the HF extrapolation has 3 parameters
and so it is necessary to use the $n=3,4$, and $5$ basis sets.  Consequently, the extrapolation error is larger for the HF energy
than for the correlation energy, mostly for molecules containing second-row atoms, as we have verified for some systems by going to the $n=6$ basis sets.
In order to partially cure this problem the cc-pV($n$+d)Z basis sets, which have one additional set of d basis functions, were introduced~\cite{DunPetWil-JCP-01}
for the second-row atoms Al through Ar.  For H, He, and first-row atoms the cc-pV$n$Z and cc-pV($n$+d)Z basis sets are identical.
Hence all the CBS energies presented in this paper use extrapolated HF energies obtained from Eq.~(\ref{HF_extrap}) but with $E^{\rm HF}_n$ replaced by $E^{\rm HF}_{n+{\rm d}}$,
where $E^{\rm HF}_{n+{\rm d}}$ are the HF energies in the cc-pV($n$+d)Z basis sets.
We find that although the cc-pV($n$+d)Z basis sets of course give lower total energies than the cc-pV$n$Z basis sets for each $n$, the estimated CBS energies are higher.
Of the systems we study, replacing the cc-pV$n$Z basis sets with the cc-pV($n$+d)Z basis sets has the largest effect for SO$_2$ and SO, reducing the atomization energies by 3.68 kcal/mol
and 0.82 kcal/mol, respectively.
The large change in the estimated CBS energy of SO$_2$ has previously been noted in Refs.~\onlinecite{WilDun-JCP-03,BauPar-CPL-95,BauRic-JPCA-98}.

To estimate the total energies in the CBS limit, we also employ the DFT-based basis-set correction recently developed in Refs.~\onlinecite{GinPraFerAssSavTou-JCP-18,LooPraSceTouGin-JPCL-19,GinSceTouLoo-JCP-19,GinSceLooTou-JCP-20}. In this scheme, the total SHCI energy in a given basis set is corrected as
\beq
\label{SHCI+pbe}
E^{\rm SHCI+PBE}_n &=& E^{\rm HF}_{n+{\rm d}} - E^{\rm HF}_n + E^{\rm SHCI}_n + \bar{E}_n^{\rm PBE}[\rho,\zeta,\mu],
\nonumber\\
\eeq
where $\bar{E}_n^{\rm PBE}[\rho,\zeta,\mu]$ is a basis-set-dependent functional of the density $\rho({\bf r})$, the spin polarization $\zeta({\bf r}) = [\rho_\uparrow({\bf r}) - \rho_\downarrow({\bf r}) ]/\rho({\bf r})$, and the local range-separation function $\mu({\bf r})$
\beq
\label{pbe_correction}
\bar{E}_n^{\rm PBE}[\rho,\zeta,\mu] = \int \rho({\bf r}) \bar{\varepsilon}_\text{c,md}^\text{sr,PBE}(\rho({\bf r}), \zeta({\bf r}), \mu({\bf r}) ) \text{d}{\bf r}. \;\;
\eeq
In Eq.~(\ref{pbe_correction}), $\bar{\varepsilon}_\text{c,md}^\text{sr,PBE}$ is the complementary short-range correlation energy per particle with multideterminant reference (md) that was constructed in Ref.~\onlinecite{LooPraSceTouGin-JPCL-19} based on the Perdew-Burke-Ernzerhof (PBE)~\cite{PerBurErn-PRL-96} correlation functional and the on-top pair density of the uniform-electron gas. The local range-separation function $\mu({\bf r})$ provides a local measure of the incompleteness of the basis set and is defined as
\beq
\label{localmu}
\mu({\bf r}) = \frac{\sqrt{\pi}}{2} W({\bf r},{\bf r}),
\eeq
where $W({\bf r},{\bf r})$ is the on-top value of the effective two-electron interaction in the basis set
\beq
\label{Weff}
W({\bf r},{\bf r}) =
\begin{cases}
	f({\bf r},{\bf r})/n_2({\bf r},{\bf r}),    & \text{if $n_2({\bf r},{\bf r}) \ne 0$,}
	\\
	\infty,                                                                                          & \text{otherwise,}
\end{cases}
\eeq
with
\begin{gather}
	\label{eq:fbasis_val}
	f({\bf r},{\bf r}) = \sum_{pq\in {\cal B}} \sum_{rstu \in {\cal A}}  \phi_p({\bf r}) \phi_q({\bf r}) V_{pq}^{rs} \Gamma_{rs}^{tu} \phi_t({\bf r}) \phi_u({\bf r}),
	\\
	\label{eq:twordm_val}
	n_2({\bf r},{\bf r})
	= \sum_{rstu \in {\cal A}} \phi_r({\bf r}) \phi_s({\bf r}) \Gamma_{rs}^{tu} \phi_t({\bf r}) \phi_u({\bf r}),
\end{gather}
where $V_{pq}^{rs}=\langle pq | rs \rangle$ are the two-electron integrals and $\Gamma_{rs}^{tu}$ is the opposite-spin two-body density matrix. Since $\mu({\bf r})$ is very weakly dependent on $\Gamma_{rs}^{tu}$, we calculate $\Gamma_{rs}^{tu}$ at the HF level only. Consistently, $\{\phi_p({\bf r})\}$ are the HF orbitals, and $\rho({\bf r})$ and $\zeta({\bf r})$ are also calculated at the HF level. Since the core electrons are frozen in SHCI, we use the frozen-core variant~\cite{LooPraSceTouGin-JPCL-19,GinSceLooTou-JCP-20} of this DFT basis-set correction which means that in Eqs.~(\ref{eq:fbasis_val}) and~(\ref{eq:twordm_val}) the sums over $r,s,t,u$ are restricted to the set of active (i.e., non-core) occupied HF orbitals ${\cal A}$. Yet, the local range-separation function $\mu({\bf r})$ probes the entire basis set through the sums over $p,q$, which run over the set of all (occupied + virtual) HF orbitals ${\cal B}$. 

For a fixed basis set, the energy functional $\bar{E}_n^{\rm PBE}[\rho,\zeta,\mu]$ provides an estimate of the energy missing in FCI to reach the CBS limit. It has the desirable property of vanishing in the CBS limit, i.e. $\bar{E}_{\rm CBS}^{\rm PBE} = 0$, and thus the DFT basis-set correction does not alter the CBS limit, i.e. $E^{\rm SHCI+PBE}_{\rm CBS} = E^{\rm SHCI}_{\rm CBS}$, but just accelerates the basis convergence. 

Based on the analysis of basis convergence in range-separated DFT~\cite{FraMusLupTou-JCP-15}, we assume an exponential basis convergence of $E^{\rm SHCI+PBE}_n$ which gives us another estimate of the CBS limit of $E^{\rm SHCI}_n$ via the extrapolation
\beq
\label{pbe_extrap}
E^{\rm SHCI+PBE}_{\rm CBS} &=& E^{\rm SHCI+PBE}_n + a \exp{(-bn)},
\eeq
using $n=3, 4, 5$. The only exceptions are Be and Cl, whose cc-pV5Z energy is higher than the cc-pVQZ energy and for which the cc-pV5Z energy is taken as the CBS energy.

\section{Computational details}
\label{Comput_details}

The HF and CCSD(T) calculations are done with PySCF~\cite{SunCha_etal_PySCF-ComMolSci-18} or MOLPRO \cite{werner2012molpro}.
The starting integrals are computed for HF orbitals. The core orbitals are kept fixed for all the subsequent steps.
Then we construct integrals in the SHCI natural orbital basis by computing and diagonalizing the
one-body density matrix and rotating the integrals in the HF basis to the natural orbital basis.
Next we use the methods discussed in Section~\ref{Orb_opt} to construct the integrals in the
optimized orbital basis. We use a fairly large value of $\epsilon_1$ (typically $2\times 10^{-4}$)
to construct the natural orbitals and the optimized orbitals.  For some systems the natural orbital
basis is reasonably close to the optimal one, but for most systems the optimized orbital bases result in
considerable gains in efficiency.  The final SHCI calculations using
the optimized orbitals employ
smaller values of $\epsilon_1$ (typically 5 values ranging from $2 \times 10^{-4}$ to $2 \times 10^{-5}$),
which are then used to extrapolate to the $\epsilon_1=0$ limit.
The system with the largest extrapolation distance, SO$_2$ in the cc-pV5Z basis, was shown as an
example in Fig.~\ref{fig:SHCI_extrap}.

The PBE-based basis-set correction described in Section \ref{basis_set} is calculated independently from the SHCI calculations using the software QUANTUM PACKAGE~\cite{GarAppGasBenFerPaqPraAssReiTouBarRenDavMalVerCafLooGinSce-JCTC-19}. 
If the HF two-body density matrix is used in Eqs.~(\ref{eq:fbasis_val}) and~(\ref{eq:twordm_val}), the basis-set correction has a computational cost of ${\cal O} (N_\text{g} N_\text{e}^2  N_{\text{orb}}^2)$ where $N_\text{g}$ is the number of real-space grid points used for numerical integration in Eq. \eqref{pbe_correction} and here $N_{\text{orb}}$ is the total number of orbitals (including core orbitals) in the basis set. The two-electron integrals in the HF orbital basis, involving up to two virtual orbitals, are also needed and the cost for doing the integral transformation to compute these is ${\cal O}(N_\text{e}^2 N_{\text{orb}}^3)$.
However, most of these integrals (aside from those involving the core orbitals) are needed for SHCI anyway.
So, the DFT-based basis-set correction does not increase the computational time of SHCI calculations appreciably.

The geometries are taken from the Supplementary Material of Ref.~\onlinecite{PetTouUmr-JCP-12},
which in turn took them from the papers cited therein. They are provided in the Supplementary Material \cite{supplementary_G2}.
The only exceptions are HCO and C$_2$H$_4$ for which we took the geometry from Ref.~\onlinecite{LooPraSceTouGin-JPCL-19}, because these geometries
gave lower CBS-extrapolated energies by approximately 1.5 mHa.
In order to compare to experimental atomization energies, the CBS SHCI energies are
corrected for
zero-point energies (ZPE), core-valence correlation (CV), scalar relativity (SR), and spin-orbit (SO) effects.
We take the corrections from the literature.  Since most of the papers do not have
all the 55 molecules we studied, we take the corrections from Refs.~\onlinecite{FelPetDix-JCP-08,FelPet-JCP-99} in that order, i.e.,
we take it from the first of these references that contains corrections for that molecule.
The source of the corrections is indicated in Table~\ref{tab:tae} next to the entry for
the zero-point energy (ZPE).
Similarly the experimental values quoted in Table~\ref{tab:tae} are taken from Refs.~\onlinecite{ATcT,ATcTb,NIST,FelPet-JCP-99} in that order.

\setlength{\tabcolsep}{8pt}
\setlength\LTcapwidth{0.75\linewidth}

\begin{table*}
\small
\begin{ruledtabular}
\caption{Deviation of SHCI and SHCI+PBE atomization energies, $D_0$, in the complete-basis-set limit, from the best available experimental energies in units of kcal/mol.  The raw SHCI and SHCI+PBE energies 
are corrected for zero-point energy (ZPE), scalar relativity (SR), spin-orbit energy (SO) and core-valence correlation (CV).
For each molecule, the ZPE, SR+SO and CV corrections are taken from Ref.~\onlinecite{FelPetDix-JCP-08} if available, and otherwise
from Ref.~\onlinecite{FelPet-JCP-99} as shown next to the ZPE correction.
The only exceptions are that the CV corrections for LiH and Li$_2$ were taken from Ref.~\onlinecite{FelPet-JCP-99} because Ref.~\onlinecite{FelPetDix-JCP-08} did not freeze the core for these systems.}
\label{tab:tae}
\begin{tabular}{lrrrdr|rr|rr}
\multicolumn{6}{c|}{} & \multicolumn{2}{c|}{SHCI} & \multicolumn{2}{c}{SHCI+PBE}\\
\cline{7-10}
         molecule & SHCI $D_e$ &    ZPE                        &  SR+SO &    \multicolumn{1}{c}{\rm{CV}} &  experiment                           & $D_0$ &  deviation &   $D_0$ &  deviation \\
\hline
              LiH &    57.71 &  -1.99  \cite{FelPetDix-JCP-08} &  -0.02 &  0.30 &       55.70               \cite{NIST} &    56.00 &         0.30 &        56.02 &             0.32 \\
              BeH &    50.23 &  -2.92  \cite{FelPetDix-JCP-08} &  -0.02 &  0.51 &       47.70  \cite{VasPetDix-JCTC-17} &    47.80 &         0.10 &        47.80 &             0.10 \\
               CH &    84.11 &  -4.04  \cite{FelPetDix-JCP-08} &  -0.08 &  0.14 &       79.97               \cite{ATcT} &    80.13 &         0.16 &        80.16 &             0.19 \\
  CH$_2$($^3B_1$) &   190.01 & -10.55  \cite{FelPetDix-JCP-08} &  -0.23 &  0.82 &      179.83               \cite{ATcT} &   180.05 &         0.22 &       179.95 &             0.12 \\
  CH$_2$($^1A_1$) &   181.12 & -10.29  \cite{FelPetDix-JCP-08} &  -0.17 &  0.39 &      170.83               \cite{ATcT} &   171.05 &         0.22 &       171.10 &             0.27 \\
           CH$_3$ &   306.93 & -18.55  \cite{FelPetDix-JCP-08} &  -0.25 &  1.07 &      289.11               \cite{ATcT} &   289.20 &         0.09 &       289.18 &             0.07 \\
           CH$_4$ &   419.25 & -27.74  \cite{FelPetDix-JCP-08} &  -0.27 &  1.26 &      392.47               \cite{ATcT} &   392.50 &         0.03 &       392.56 &             0.09 \\
               NH &    83.09 &  -4.64  \cite{FelPetDix-JCP-08} &  -0.07 &  0.11 &       78.36               \cite{ATcT} &    78.49 &         0.13 &        78.55 &             0.19 \\
           NH$_2$ &   182.50 & -11.84  \cite{FelPetDix-JCP-08} &   0.08 &  0.32 &      170.59               \cite{ATcT} &   171.06 &         0.47 &       171.10 &             0.51 \\
           NH$_3$ &   297.91 & -21.33  \cite{FelPetDix-JCP-08} &  -0.25 &  0.65 &      276.59               \cite{ATcT} &   276.98 &         0.39 &       276.97 &             0.38 \\
               OH &   107.26 &  -5.29  \cite{FelPetDix-JCP-08} &  -0.24 &  0.14 &      101.73               \cite{ATcT} &   101.87 &         0.14 &       101.81 &             0.08 \\
           H$_2$O &   233.01 & -13.26  \cite{FelPetDix-JCP-08} &  -0.49 &  0.38 &      219.37               \cite{ATcT} &   219.64 &         0.27 &       219.51 &             0.14 \\
               HF &   141.76 &  -5.86  \cite{FelPetDix-JCP-08} &  -0.58 &  0.17 &      135.27               \cite{ATcT} &   135.49 &         0.22 &       135.37 &             0.10 \\
 SiH$_2$($^1A_1$) &   153.90 &  -7.30     \cite{FelPet-JCP-99} &  -0.60 &  0.00 &      144.10               \cite{NIST} &   146.00 &         1.90 &       146.05 &             1.95 \\
 SiH$_2$($^3B_1$) &   133.31 &  -7.50     \cite{FelPet-JCP-99} &  -0.80 & -0.50 &      123.40      \cite{FelPet-JCP-99} &   124.51 &         1.11 &       124.42 &             1.02 \\
          SiH$_3$ &   228.22 & -13.20     \cite{FelPet-JCP-99} &  -0.80 & -0.20 &      212.20               \cite{NIST} &   214.02 &         1.82 &       214.02 &             1.82 \\
          SiH$_4$ &   324.80 & -19.40     \cite{FelPet-JCP-99} &  -1.00 & -0.20 &      302.60               \cite{NIST} &   304.20 &         1.60 &       304.27 &             1.67 \\
           PH$_2$ &   154.24 &  -8.40     \cite{FelPet-JCP-99} &  -0.20 &  0.30 &      144.70      \cite{FelPet-JCP-99} &   145.94 &         1.24 &       145.96 &             1.26 \\
           PH$_3$ &   241.91 & -14.44  \cite{FelPetDix-JCP-08} &  -0.44 &  0.33 &      227.10               \cite{NIST} &   227.36 &         0.26 &       227.36 &             0.26 \\
           H$_2$S &   183.63 &  -9.40  \cite{FelPetDix-JCP-08} &  -0.93 &  0.24 &      173.20               \cite{NIST} &   173.54 &         0.34 &       173.41 &             0.21 \\
              HCl &   107.41 &  -4.24     \cite{FelPet-JCP-99} &  -1.00 &  0.30 &      102.21               \cite{ATcT} &   102.47 &         0.26 &       102.30 &             0.09 \\
           Li$_2$ &    24.14 &  -0.50  \cite{FelPetDix-JCP-08} &   0.00 &  0.20 &       23.90               \cite{NIST} &    23.84 &        -0.06 &        23.84 &            -0.06 \\
              LiF &   138.15 &  -1.30     \cite{FelPet-JCP-99} &  -0.60 &  0.90 &      137.60               \cite{NIST} &   137.15 &        -0.45 &       137.34 &            -0.26 \\
       C$_2$H$_2$ &   403.16 & -16.50  \cite{FelPetDix-JCP-08} &  -0.46 &  2.47 &      388.64               \cite{ATcT} &   388.67 &         0.03 &       388.84 &             0.20 \\
       C$_2$H$_4$ &   561.72 & -31.66  \cite{FelPetDix-JCP-08} &  -0.50 &  2.36 &      532.04               \cite{ATcT} &   531.92 &        -0.12 &       532.09 &             0.05 \\
       C$_2$H$_6$ &   711.36 & -46.23  \cite{FelPetDix-JCP-08} &  -0.56 &  2.42 &      666.19               \cite{ATcT} &   666.99 &         0.80 &       666.97 &             0.78 \\
               CN &   180.24 &  -2.95  \cite{FelPetDix-JCP-08} &  -0.24 &  1.10 &      178.12               \cite{ATcT} &   178.15 &         0.03 &       178.58 &             0.46 \\
              HCN &   311.91 &  -9.95  \cite{FelPetDix-JCP-08} &  -0.31 &  1.67 &      303.14               \cite{ATcT} &   303.32 &         0.18 &       303.76 &             0.62 \\
               CO &   258.61 &  -3.09  \cite{FelPetDix-JCP-08} &  -0.46 &  0.95 &      256.23               \cite{ATcT} &   256.01 &        -0.22 &       256.47 &             0.24 \\
              HCO &   278.10 &  -8.09  \cite{FelPetDix-JCP-08} &  -0.59 &  1.16 &      270.76               \cite{ATcT} &   270.58 &        -0.18 &       270.92 &             0.16 \\
          H$_2$CO &   373.42 & -16.52  \cite{FelPetDix-JCP-08} &  -0.65 &  1.30 &      357.48               \cite{ATcT} &   357.55 &         0.07 &       357.88 &             0.40 \\
         H$_3$COH &   512.44 & -31.72     \cite{FelPet-JCP-99} &  -0.80 &  1.50 &      480.97               \cite{ATcT} &   481.42 &         0.45 &       481.52 &             0.55 \\
            N$_2$ &   227.66 &  -3.36  \cite{FelPetDix-JCP-08} &  -0.14 &  0.80 &      224.94               \cite{ATcT} &   224.96 &         0.02 &       225.62 &             0.68 \\
       N$_2$H$_4$ &   438.61 & -32.68  \cite{FelPetDix-JCP-08} &  -0.51 &  1.14 &      404.81               \cite{ATcT} &   406.56 &         1.75 &       406.60 &             1.79 \\
               NO &   152.33 &  -2.71  \cite{FelPetDix-JCP-08} &  -0.23 &  0.42 &      149.81               \cite{ATcT} &   149.81 &         0.00 &       150.23 &             0.42 \\
            O$_2$ &   120.50 &  -2.25  \cite{FelPetDix-JCP-08} &  -0.62 &  0.24 &      117.99               \cite{ATcT} &   117.87 &        -0.12 &       117.95 &            -0.04 \\
       H$_2$O$_2$ &   269.21 & -16.44  \cite{FelPetDix-JCP-08} &  -0.82 &  0.36 &      252.21               \cite{ATcT} &   252.31 &         0.10 &       252.33 &             0.12 \\
            F$_2$ &    39.09 &  -1.30  \cite{FelPetDix-JCP-08} &  -0.79 & -0.11 &       36.93               \cite{ATcT} &    36.89 &        -0.04 &        36.93 &             0.00 \\
           CO$_2$ &   388.19 &  -7.24  \cite{FelPetDix-JCP-08} &  -1.01 &  1.77 &      381.98               \cite{ATcT} &   381.71 &        -0.27 &       382.46 &             0.48 \\
           Na$_2$ &    16.74 &  -0.20     \cite{FelPet-JCP-99} &   0.00 &  0.30 &       17.00               \cite{NIST} &    16.84 &        -0.16 &        16.85 &            -0.15 \\
           Si$_2$ &    76.66 &  -0.73  \cite{FelPetDix-JCP-08} &  -1.01 &  0.13 &       74.40               \cite{NIST} &    75.05 &         0.65 &        75.03 &             0.63 \\
            P$_2$ &   116.66 &  -1.11  \cite{FelPetDix-JCP-08} &  -0.25 &  0.77 &      116.00               \cite{NIST} &   116.07 &         0.07 &       116.29 &             0.29 \\
            S$_2$ &   103.95 &  -1.04  \cite{FelPetDix-JCP-08} &  -1.40 &  0.34 &      100.80               \cite{NIST} &   101.85 &         1.05 &       101.51 &             0.71 \\
           Cl$_2$ &    59.92 &  -0.80  \cite{FelPetDix-JCP-08} &  -1.82 & -0.13 &       57.18               \cite{ATcT} &    57.17 &        -0.01 &        56.75 &            -0.43 \\
             NaCl &   100.03 &  -0.50     \cite{FelPet-JCP-99} &  -1.10 & -1.20 &       97.40               \cite{NIST} &    97.23 &        -0.17 &        96.85 &            -0.55 \\
              SiO &   192.01 &  -1.78  \cite{FelPetDix-JCP-08} &  -0.90 &  0.95 &      189.80               \cite{NIST} &   190.28 &         0.48 &       190.53 &             0.73 \\
               CS &   171.55 &  -1.83  \cite{FelPetDix-JCP-08} &  -0.80 &  0.75 &      170.40               \cite{NIST} &   169.67 &        -0.73 &       169.67 &            -0.73 \\
               SO &   126.15 &  -1.63  \cite{FelPetDix-JCP-08} &  -1.09 &  0.41 &      123.50               \cite{NIST} &   123.84 &         0.34 &       123.67 &             0.17 \\
              ClO &    65.58 &  -1.22  \cite{FelPetDix-JCP-08} &  -0.81 &  0.06 &       63.42               \cite{ATcT} &    63.61 &         0.19 &        63.07 &            -0.35 \\
              ClF &    62.95 &  -1.12  \cite{FelPetDix-JCP-08} &  -1.39 & -0.10 &       60.35               \cite{ATcT} &    60.34 &        -0.01 &        59.99 &            -0.36 \\
      Si$_2$H$_6$ &   535.40 & -30.50     \cite{FelPet-JCP-99} &  -2.00 &  0.00 &      500.10      \cite{FelPet-JCP-99} &   502.90 &         2.80 &       503.34 &             3.24 \\
         CH$_3$Cl &   395.06 & -23.19     \cite{FelPet-JCP-99} &  -1.40 &  1.20 &      371.35               \cite{ATcT} &   371.67 &         0.32 &       371.53 &             0.18 \\
         H$_3$CSH &   474.48 & -28.60     \cite{FelPet-JCP-99} &  -1.20 &  1.50 &      445.10               \cite{NIST} &   446.18 &         1.08 &       445.91 &             0.81 \\
             HOCl &   166.62 &  -8.18     \cite{FelPet-JCP-99} &  -1.50 &  0.40 &      156.88               \cite{ATcT} &   157.34 &         0.46 &       156.93 &             0.05 \\
           SO$_2$ &   260.36 &  -4.38  \cite{FelPetDix-JCP-08} &  -1.79 &  0.92 &      254.46              \cite{ATcTb} &   255.11 &         0.65 &       255.00 &             0.54 \\
\end{tabular}
\end{ruledtabular}
\end{table*}

\section{Results}
\label{Results}

\subsection{Accuracy of CCSD(T)}

\begin{figure*}[htb]
 \centerline{\includegraphics[width=1.15\textwidth]{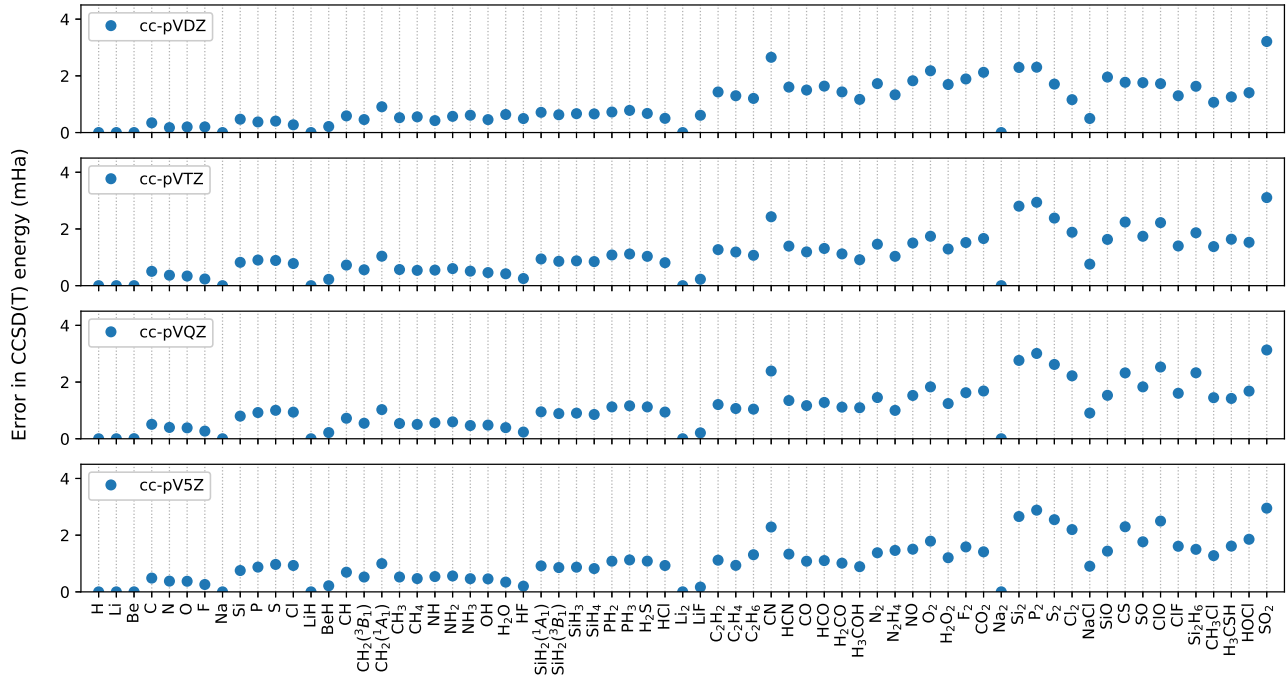}}
 \caption{
 The error in the CCSD(T) total energies obtained by comparison to the SHCI total energies.
 The CCSD(T) errors are of course zero for systems with one or two valence electrons, and they are positive
 in all other cases.
 The errors for each system are very similar for the various basis sets, especially for the larger basis sets.}
 \label{fig:ccsdt}
\end{figure*}

\begin{figure*}[htb]
	\centerline{\includegraphics[width=1.15\textwidth]{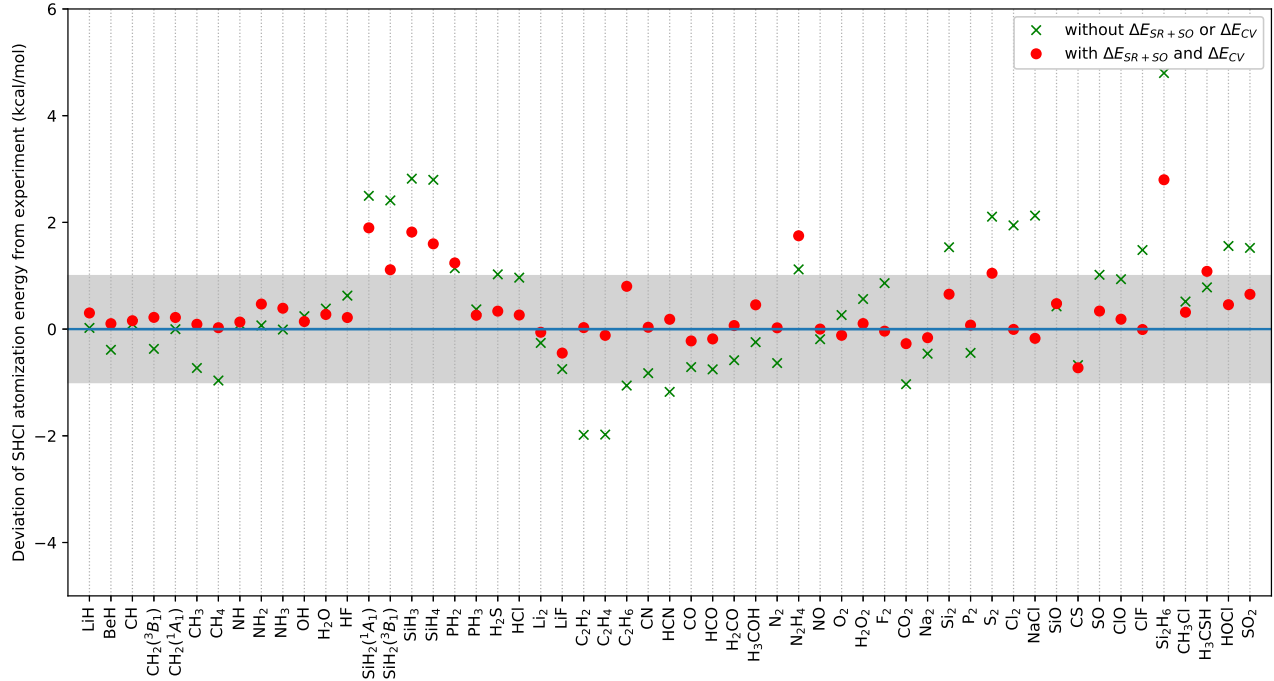}}
	\caption{The comparison of SHCI atomization energies in the extrapolated complete-basis-set limit with experiment, with (red dots) and without (green crosses) scalar relativistic and spin-orbit (SR+SO) corrections and core-valence (CV) corrections. Both sets of points include zero-point energy (ZPE) corrections. Systems for which red dots fall in the shaded region are considered to have reached chemical accuracy (1 kcal/mol).}
	\label{fig:atomization}
\end{figure*}

\begin{figure*}[htb]
	\centerline{\includegraphics[width=1.15\textwidth]{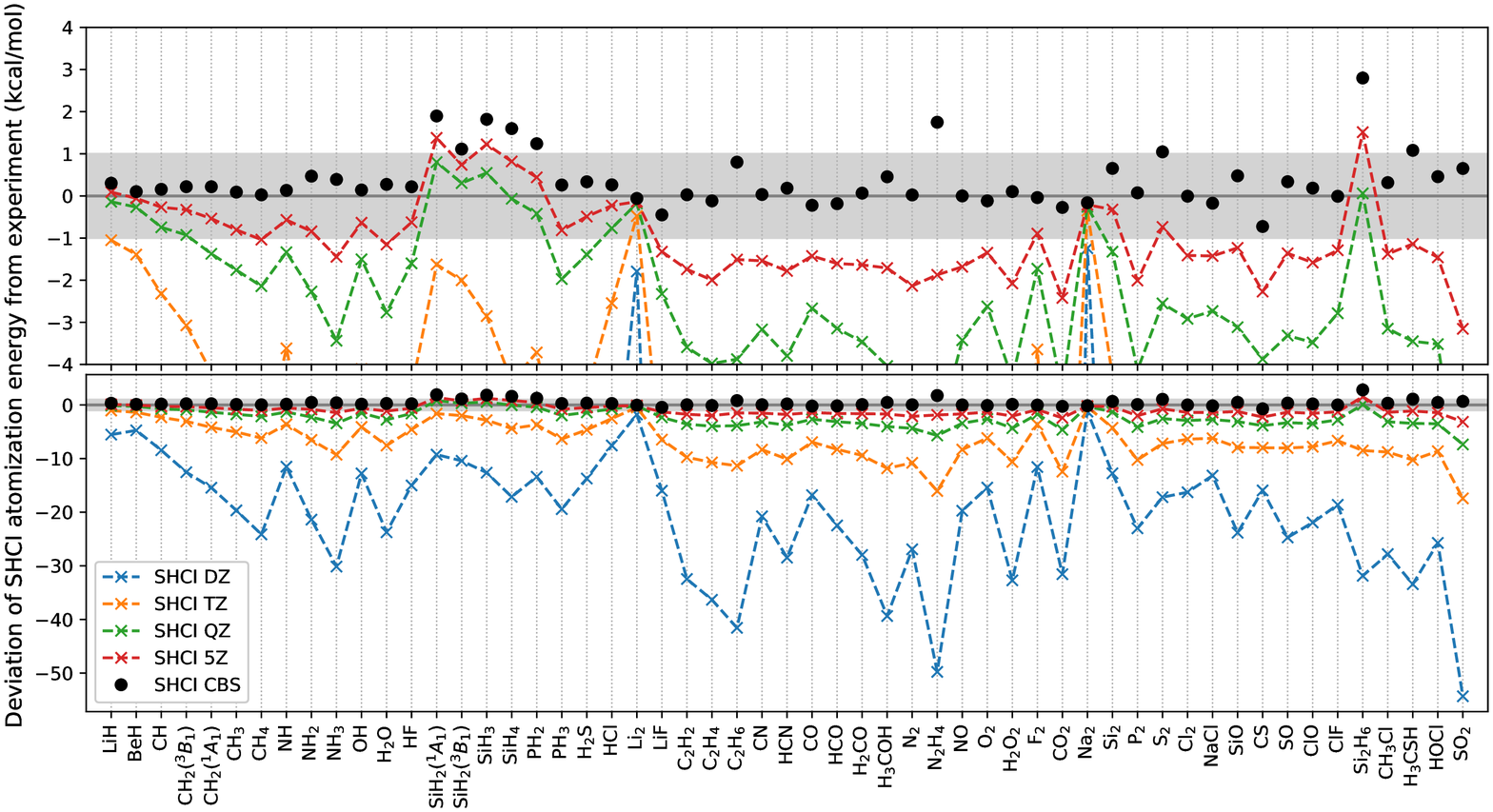}}
	\caption{The comparison of SHCI atomization energies with experiment in the individual basis sets and in the extrapolated complete-basis-set limit. The top panel is a blowup of the top portion of the bottom panel. The shaded region indicates chemical accuracy (1 kcal/mol).}
	\label{fig:atomization_shci}
\end{figure*}

\begin{figure*}[htb]
	\centerline{\includegraphics[width=1.15\textwidth]{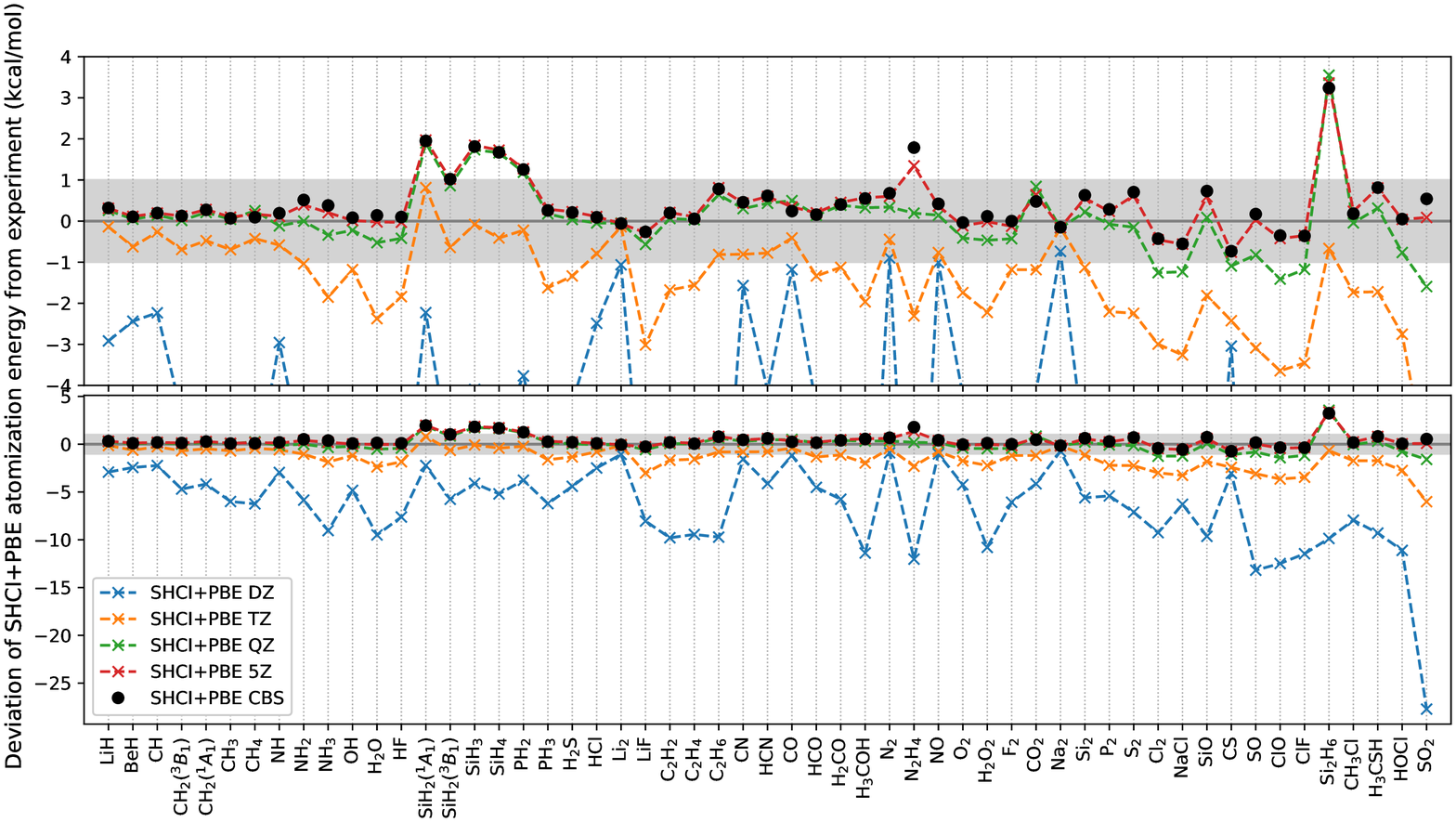}}
	\caption{Same as Fig.~\ref{fig:atomization_shci} but with the PBE-based basis set correction applied.
        The extrapolation distances are much reduced compared to Fig.~\ref{fig:atomization_shci}.}
	\label{fig:atomization_shci_pbe}
\end{figure*}

We have computed the total energies for each of the 55 molecules and their 12 constituent atoms in the four basis sets mentioned in
Section~\ref{basis_set}.
The accuracy of these energies should be considerably better than 1 mHa, as discussed later in this section.
These energies are provided in CSV files in the Supplementary Material~\cite{supplementary_G2}
and can serve as a reference for other approximate methods.
In particular, we have used it to test the accuracy of CCSD(T).
None of the 67 systems studied is strongly correlated, so one would expect the CCSD(T) energies to
be reasonably accurate.
This is in fact the case, as can be seen from Fig. \ref{fig:ccsdt}, which shows the deviation of the CCSD(T)
total energies from the SHCI total energies.
CCSD(T) deviates from SHCI by 1-2 mHa for the lighter systems and
3-4 mHa for the heavier ones.
For systems with two or fewer valence electrons, the two methods agree exactly as they must, and
for all the systems with more electrons, CCSD(T) underestimates the correlation energy.
The mean absolute deviation (MAD) is roughly independent of the basis size, being 0.99, 1.06, 1.09, and 1.05 mHa, respectively,
for the four basis sets.
The pattern of the errors is very similar for the four basis sets.
Although the absolute value of the correlation energy grows with the size of the basis set by a few tens of percent
going from cc-pVDZ to cc-pV5Z basis sets, the error that CCSD(T) makes does not grow in proportion.

The same set of molecules have also recently been computed by another SCI+PT method~\cite{TubFreLevHaiHeaWha-JCTC-20}.
In their calculation they correlate all the electrons, so the energies they obtain are not directly comparable to ours.
They employ only the cc-pVDZ and cc-pVTZ basis sets so they cannot extrapolate to the CBS limit.
Further, they employ at the most only $10^6$ determinants, whereas we employ a few times $10^8$ determinants for
the larger molecules and basis sets.  Consequently when they compare to CCSD(T) energies, they find two systems for
the cc-pVDZ basis set and several systems for the cc-pVTZ basis set where their energies are higher than those from CCSD(T).
In contrast, as shown in Fig.~\ref{fig:ccsdt}, we find that our SHCI energies are always lower than CCSD(T) energies
and further that the pattern of the energy differences is very similar for the various basis sets.

\subsection{Atomization energies}
Table~\ref{tab:tae} shows the difference between the SHCI total energies for the molecules and their constituent
atoms, extrapolated to the CBS limit according to Eqs. (\ref{HF_extrap}) and (\ref{corr_extrap}).
It also shows the ZPE, SR+SO, and CV corrections taken from the literature and the final prediction for the
SHCI atomization energy, $D_0$, and how much it differs from the best available experimental values.
The difference between the SHCI $D_0$ and experiment is also plotted in Fig. \ref{fig:atomization},
both before and after the corrections are applied.

There are 3 possible sources of discrepancy between the calculated and the experimental atomization energies: (1) The extrapolation to the CBS limit may not be accurate; (2) the literature values of the ZPE, SR+SO, and CV corrections may not be accurate; 3) the experimental values have errors.
It seems likely, as discussed below, that all three of these play a role for some of the systems.

We show in Fig. \ref{fig:atomization_shci} the convergence of the atomization energies with basis size. The SHCI atomization energies in fact have two extrapolation errors.
The first and more benign error comes from extrapolating SHCI total energies for each basis set to the FCI limit, i.e.,
$\epsilon_1\rightarrow0$. This error can be reduced by employing smaller $\epsilon_1$ and/or 
using better optimized orbitals. For the four basis sets $n$ = 2 (D), 3 (T), 4 (Q), and 5, the largest extrapolation distances in the total energy of these
55 molecules and 12 atoms are 0.97, 2.36, 3.34, and 2.90 mHa, respectively.
\footnote{The extrapolation distance depends on the value of $\epsilon_1$ in Eq.~\ref{HCI_criterion} and
on how well the orbitals are optimized to improve the convergence of the energy.}
Assuming that the extrapolated energies are in error by no more than a fifth of the extrapolation distance,
all these energies should be accurate to considerably better than 1 mHa.
Further, the typical extrapolation
distances are much smaller, especially for the lighter systems: the median distances for the four basis sets are
2.92, 14.4, 56.4, and 77.0 $\mu$Ha, respectively.
The second source of error comes from extrapolation to the CBS
limit, using Eqs. (\ref{HF_extrap}) and (\ref{corr_extrap}), and is less under control. For these 67 systems, the maximum and median CBS extrapolation
distances are 21.8 and 6.47 mHa, respectively.
This CBS extrapolation error is likely to be an important error for those systems where the
extrapolation distance (the energy difference between the black dots and red crosses in Fig.~\ref{fig:atomization_shci})
is large.

To further study the magnitude of the CBS extrapolation error, we add the PBE-based basis-set correction discussed in Sec.~\ref{basis_set}
to the SHCI energies for each basis set [see Eqs.~(\ref{SHCI+pbe}) and~(\ref{pbe_correction})] and then
extrapolate the corrected energies to the CBS limit according to Eq.~(\ref{pbe_extrap}),
which gives us an alternative way to estimate the CBS limit of the SHCI energies. The PBE-based corrections can also be found in the Supplemental Material~\cite{supplementary_G2}.
It is apparent from Table~\ref{tab:tae} that the deviations of the SHCI and the SHCI+PBE energies from experiment
are strongly correlated, thereby giving us a reasonable measure of confidence in our two extrapolations
as well as an estimate of the extrapolation errors.
Fig. \ref{fig:atomization_shci_pbe} shows the same information as Fig.~\ref{fig:atomization_shci} after the PBE-based basis-set correction has been included.
As summarized in Table \ref{tab:summary_statistics}, for each basis set the MAD from experiment decreases by about a factor of 3 compared to that without the basis-set correction.
\footnote{To avoid confusion, we note that in Ref.~\onlinecite{LooPraSceTouGin-JPCL-19} it was found that
CCSD(T)+PBE had a MAD of only 1.96, 0.85, and 0.31 kcal/mol with respect to the CCSD(T) CBS limit for the cc-pVDZ, cc-pVTZ, and cc-pVQZ basis sets, respectively.  These considerably smaller values compared to those in
Table~\ref{tab:summary_statistics} are the result of a one-body basis-set correction that was always included by
adding the cc-pV5Z HF energy to the CCSD(T) correlation energies for the different basis sets.
Of course one could do the same for the SHCI energies in the current paper.}
In particular, SHCI+PBE gives a MAD of only 0.55 kcal/mol already with the cc-pVQZ basis set.
The cc-pV5Z basis set has a MAD of only 0.49 kcal/mol.
Applying the CBS extrapolation to SHCI+PBE gives a somewhat larger MAD from experiment of 0.51 kcal/mol, as the computed atomization energies are too small for the smaller basis sets but increase with increasing
basis size and 
for the majority of the molecules the computed CBS atomization energies are larger than experiment.

As seen from Figs.~\ref{fig:atomization_shci} and \ref{fig:atomization_shci_pbe} the predicted CBS atomization
energy of Si$_2$H$_6$ is more than 3 kcal/mol larger than experiment.  However, even the
$n=5$
value is larger than experiment, so the discrepancy cannot be attributed to
an inaccurate CBS extrapolation, but instead to either inaccurate ZPE, SR+SO, and CV corrections, or,
to errors in the experimental value.
The ZPE correction for Si$_2$H$_6$ is quite large, -30.50 mHa, so even a small fractional error
in its estimate could account for the discrepancy in the atomization energy.
In fact, these statements hold for all seven molecules in Fig.~\ref{fig:atomization_shci_pbe} that
have cc-pV5Z atomization energies that are larger than experiment by more than 1 kcal/mol.
Note that there are several systems for which the atomization energies are overestimated in Figs.~\ref{fig:atomization_shci} and \ref{fig:atomization_shci_pbe}
by more than 1 kcal/mol, but none for which they are underestimated by more than 1 kcal/mol.

The majority of the deviations fall below 1 kcal/mol, reaching chemical accuracy as can be seen in Table~\ref{tab:tae}
and Figs.~\ref{fig:atomization_shci} and \ref{fig:atomization_shci_pbe}.
As regards those where the deviations are larger than 1 kcal/mol it should also be kept in mind that
that in addition to the uncertainties in the corrections, especially the ZPE correction,
the experimental values may also be inaccurate, particularly for those atomization energies that are not available
from the ATcT database~\cite{ATcT}.
For example, for PH$_2$ the two available experimental values differ by 4.5 kcal/mol and
our computed value differs by +1.5 kcal/mol from Ref.~\onlinecite{FelPet-JCP-99}
and -3.0 kcal/mol from Ref.~\onlinecite{NIST}.
For the molecules in the ATcT database the MAD is only 0.24 kcal/mol before the PBE-based basis set correction is applied
and 0.32 kcal/mol after it is applied. 

\begin{table}
  \caption{Summary statistics of deviations from experimental atomization energies for the 55 molecules.
    For each of the basis sets (but not for the CBS limit) the inclusion of the PBE-based basis-set correction reduces the MAD by about
    a factor of 3.
    MAD: mean absolute deviation. MAX: maximum absolute deviation. Units: kcal/mol.
  }
  \begin{ruledtabular}
    \begin{tabular}{lrr}
      Method & MAD & MAX \\
      \hline
      SHCI cc-pVDZ &  20.77 & 54.32 \\
      SHCI cc-pVTZ &   6.83 & 17.43 \\
      SHCI cc-pVQZ &   2.47 &  7.38 \\
      SHCI cc-pV5Z &   1.20 &  3.15 \\
      SHCI CBS     &   0.46 &  2.80 \\
      \hline
      SHCI+PBE cc-pVDZ &   6.52 & 27.72 \\
      SHCI+PBE cc-pVTZ &   1.47 &  6.02 \\
      SHCI+PBE cc-pVQZ &   0.55 &  3.55 \\
      SHCI+PBE cc-pV5Z &   0.49 &  3.36 \\
      SHCI+PBE CBS     &   0.51 &  3.24 \\
    \end{tabular}
  \end{ruledtabular}
  \label{tab:summary_statistics}
\end{table}

Compared to other methods, our MAD of 0.46 kcal/mol is significantly less than the MAD of 1.2 to 3.2 kcal/mol
obtained in various QMC studies~\cite{Gro-JCP-02,NemTowNee-JCP-10,PetTouUmr-JCP-12}.
Diffusion Monte Carlo works directly in the CBS limit, but the fixed-node approximation is the dominant error.
Using trial wave functions with Slater determinants chosen from an SCI method, it should be easily possible
to reduce considerably the fixed-node error as demonstrated in Refs.~\onlinecite{GinSceCaf-CJC-13,GinAssTou-MP-16,DasMorSceFil-JCTC-18}.
Our MAD is comparable to results reported from composite coupled-cluster-based methods
\cite{FelPet-JCP-99,MarOli-JCP-99,HauKlo-JCP-12}.
The HEAT studies performed all-electron calculations using the coupled-cluster method with up to quadruple excitations
on a somewhat different set of molecules consisting solely of first-row elements\cite{TajSzaCsaKalGauValFloQueSta-JCP-04}.
Unfortunately, none of the molecules for which we have discrepancies of more than 1 kcal/mol were included.
For the 19 molecules also present in the G2 set, the MAD of HEAT, SHCI, and SHCI+PBE are 0.07, 0.16, and 0.27 kcal/mol, respectively.
It should be noted that HEAT is a composite quantum chemistry method, and for the lower levels of theory
it employs larger basis sets than those we used, thereby significantly reducing the CBS extrapolation error.

\section{Conclusion and outlook}
\label{conclusion}
The SHCI method enables the calculation of essentially exact energies within basis sets up to cc-pV5Z of all the molecules in the G2 set. After extrapolation to the CBS limit and addition of ZPE, SR+SO and CV corrections, the MAD from the experimental atomization energies
is only about 0.5 kcal/mol. However, depending on whether we use the PBE-based basis-set corrections or not,
there there are 7 or 9 molecules where the computed atomization energy
is more than 1 kcal/mol larger than experiment (and none for which it is more than 1 kcal/mol smaller than experiment).
These differences are mostly due to a combination of errors in the various corrections applied and in the experiments
rather than lack of convergence of the SHCI energies to the FCI energies.
With additional computational effort it would be possible to reduce the uncertainties in the computed energies.
First, instead of adding on a CV energy correction, one could use the cc-pwCVnZ basis sets to include the correlation contribution from the core electrons.  This could also make the basis-set extrapolation more reliable.
Although this entails a large increase in the Hilbert space, the increase
in the computational cost of the SHCI is not prohibitive because relatively few of the core excitations have
a large amplitude.
Second, relativistic effects could also be included within the SHCI method, as has already been demonstrated~\cite{MusSha-JCTC-18}.
Third, the computation of the ZPE correction would require calculating derivatives with respect to the nuclear
coordinates.  This could also be done, but would be the most computationally expensive part of
the calculation. Fourth, the CBS extrapolation could be improved either by employing better basis sets or using a better DFT-based
basis-set correction that employs the SHCI rather than the HF density matrix.
With these improvements, the computed energies could be sufficiently accurate to reliably pinpoint errors in experimental values of atomization energies.

\begin{acknowledgements}
This work was supported in part by the AFOSR under grant FA9550-18-1-0095.
Y.Y. acknowledges support from the Molecular Sciences Software Institute, funded by U.S. National Science Foundation grant ACI-1547580.
Some of the computations were performed at the Bridges cluster at the Pittsburgh Supercomputing Center supported by NSF grant ACI-1445606.
We thank Pierre-Fran\c{c}ois Loos for valuable comments on the manuscript and helping us converge the HF calculation
of Si$_2$ to the correct $^3\Sigma_g^-$ ground state, and one of the referees for suggesting that
we use the cc-pV($n$+d)Z basis sets to improve the basis-set convergence.
\end{acknowledgements}

\section*{Data availability}
The data that support the findings of this study are available within the article and the supplementary material of
the arXiv version of this paper~\cite{YaoGinLiTouUmr-ARX-20}.

\bibliographystyle{apsrev4-1}
\bibliography{all}

\end{document}